\pdfoutput = 1
\documentclass[apj,numberedappendix]{emulateapj}

\usepackage{epsfig}
\usepackage{amsmath}
\usepackage{enumerate}
\usepackage{mathrsfs}

\usepackage{natbib}
\usepackage{hyperref}

\usepackage{ulem}
\usepackage[squaren, Gray, cdot]{SIunits}

\usepackage{color}

\newbox\grsign \setbox\grsign=\hbox{$>$} \newdimen\grdimen \grdimen=\ht\grsign
\newbox\simlessbox \newbox\simgreatbox \newbox\simpropbox
\setbox\simgreatbox=\hbox{\raise.5ex\hbox{$>$}\llap
     {\lower.5ex\hbox{$\sim$}}}\ht1=\grdimen\dp1=0pt
\setbox\simlessbox=\hbox{\raise.5ex\hbox{$<$}\llap
     {\lower.5ex\hbox{$\sim$}}}\ht2=\grdimen\dp2=0pt
\setbox\simpropbox=\hbox{\raise.5ex\hbox{$\propto$}\llap
     {\lower.5ex\hbox{$\sim$}}}\ht2=\grdimen\dp2=0pt
\def\simgt{\mathrel{\copy\simgreatbox}}

\def\beq{\begin{equation}}
\def\eeq{\end{equation}}

\renewcommand{\vec}[1]{\pmb{#1}}
\begin{document}

\title{Precessing flaring magnetar as a source of repeating FRB~180916.J0158+65}

\author{Yuri Levin$^{1,2,3}$, Andrei M. Beloborodov$^{1,4}$, and Ashley Bransgrove$^1$.}
\affil{$^1$Physics Department and Columbia Astrophysics Laboratory, Columbia University, 
538 West 120th Street, New York, NY 10027}
\affil{$^2$Center for Computational Astrophysics, Flatiron Institute, 162 5th Ave, NY10011}
\affil{$^3$School of Physics and Astronomy, Monash University, Clayton VIC 3800, Australia}
\affil{$^4$Max Planck Institute for Astrophysics, Karl-Schwarzschild-Str. 1, D-85741, Garching, Germany}

\begin{abstract}

Recently, CHIME detected periodicity in the bursting rate of the repeating FRB~180916.J0158+65. In a popular class of models, the fast radio bursts (FRBs) are created by giant magnetic flares of a hyper-active magnetar driven by fast ambipolar diffusion in the core. We point out that in this scenario the magnetar is expected to precess freely with a period of  hours  to weeks. The internal magnetic field $B\sim 10^{16}$~G deforms the star, and 
magnetic flares induce sudden changes in magnetic stresses. The resulting torques and displacements of the principal axes of inertia 
are capable of pumping a significant amplitude of precession.
The anisotropy of the flaring FRB activity, combined with precession, implies a strong periodic modulation of the visible bursting rate.
The ultra-strong field invoked in the magnetar model provides: (1) energy for the frequent giant flares, (2) the high rate of ambipolar diffusion, releasing the magnetic energy  on the timescale $\sim 10^9$~s, (3) the core temperature $T\approx 10^9$~K, likely above the critical temperature for neutron superfluidity, 
(4) strong magnetospheric torques, which efficiently spin down the star, 
and (5) deformation with ellipticity $\epsilon\simgt 10^{-6}$, much greater than the rotational deformation.
These conditions result in a precession with negligible viscous damping, and can explain the observed 16-day period in FRB~180916.J0158+65. {The 
increase of
precession period 
due to the 
magnetar
spindown 
should become measurable in the near future}.
\medskip
\end{abstract}


\section{Introduction}
The Canadian Hydrogen Intensity Mapping Experiment (CHIME) is revolutionizing studies of Fast Radio Bursts (FRBs). Over the past year, nine new repeating FRBs have been found \citep{2020arXiv200103595F}, and one of them has been localized to a nearby spiral galaxy  \citep{2020Natur.577..190M}. This is a major increase in observational information on repeating FRBs, of which until last year, only one was known and well studied  \citep{2016Natur.531..202S, 2017Natur.541...58C}. Recently, \cite{2020arXiv200110275T} 
reported a strong periodicity of $16.35$ days in the rate of bursts from a repeating source FRB 180916.J0158+65. The bursts were observed only during a particular $\sim 5$-days long phase window of the whole $16.35$-days period, with several ($0$ to $5$) bursts arriving during each cycle. The reader is urged to inspect the striking Figure 2 of the discovery paper.

The nature of the periodicity holds an important clue to the nature of repeating FRBs. \cite{2020arXiv200110275T} suggest that the periodicity is caused either by an interaction with a companion, or by precession of a neutron star that generates the bursts. They point out that in principle the periodicity could also be caused by the spin of the neutron star, as was previously suggested by \cite{2019arXiv190900004M}, but discount this by noting that $16$-day period would be unexpectedly slow for a young object. In this letter we explore free precession as the origin of periodicity.\footnote{We also refer readers to a recent preprint by \cite{Lyutikov2020} which explores a scenario with a companion. These authors  note that geodetic precession is unlikely to produce the required periodicity. We emphasize that in our model the precession is free and does not require the presence of any companion.}  

\section{Free precession of a magnetar}

\subsection{Appearance of a precessing FRB-producing neutron star}

We focus on a class of scenarios, in which the bursts are powered by giant flares of magnetars
\citep{2013arXiv1307.4924P, 2014MNRAS.442L...9L, 2017ApJ...843L..26B,  2019arXiv190807743B, 2019MNRAS.485.4091M, 2019arXiv191105765M, Lyubarsky2020}. In these models, the hyper-activity of repeating FRBs results from fast ambipolar diffusion of the magnetic field in the magnetar core, on the timescale $\sim 10^9$~s \citep{2016ApJ...833..261B,2017ApJ...843L..26B}.

The location of coherent radio wave emission in these scenarios is the topic of current debates. FRB production inside the magnetosphere is discussed as one possibility (e.g., \citealp{2016ApJ...826..226K, 2018MNRAS.477.2470L, 2019arXiv190103260L}). Another possibility is the emission from a much larger radius outside the light cylinder of the rotating neutron star. The magnetospheric flares eject magnetically dominated plasmoids \citep{2013ApJ...774...92P}, which expand, accelerate, and flatten into a  pancake-like shape as they fly away from the star \citep{2010PhRvE..82e6305L, 2011MNRAS.411.1323G}. In the blast wave model of Beloborodov (2017, 2019), this magnetic ``pancake'' drives a shock into the magnetar wind, which generates coherent radio emission via a shock maser mechanism. The pancake occupies a significant solid angle \citep{2020arXiv200106037M}, and its emission has extreme Doppler beaming so that  observers outside that solid angle are unable to detect an FRB. We  emphasize, however, that what follows does not depend on the details of the emission  scenario, and will be equally applicable to any model in which FRBs are emitted by a magnetar with anisotropic bursting activity. 

The magnetar model relies on a superstrong  magnetic field inside the neutron star, $B\sim 10^{16}$~G.
It gives both a large energy budget, sufficent to power the observed FRBs with efficiency as low as $10^{-6}$, and the high rate of ambipolar diffusion which leads to frequent giant flares of the young magnetar. This field also deforms the magnetar, giving it an ellipticity of 
\begin{equation}
  \epsilon=k\times 10^{-4}\left({B_{\rm int}\over 10^{16}\,\hbox{G}}\right)^2,
  \end{equation}
where $B_{\rm int}$ is the characteristic internal magnetic field and $k$ is a numerical coefficient. The maximum $k\approx 1$ would be approached if the field is fully coherent and purely toroidal \citep{1969ApJ...157.1395O, 2002PhRvD..66h4025C}. There are not many explicit computations of deformations from magnetic fields 
with more realistic configurations. \cite{2015MNRAS.447.3475M} demonstrate that an internal poloidal field can dramatically decrease the ellipticity (see, e.g., Figure 5 in their paper). The value of $k$ is also reduced if the field is tangled, as expected if the field was generated immediately after the magnetar birth when its cooling involved convection. Therefore, $k\ll 1$ is expected. 

The spindown of the magnetar is controlled by its magnetic dipole moment $\mu$.
The dipole field component $B_{\rm dip}\equiv \mu/R^3$ (where $R$ is the radius of the star) can be much smaller than $B_{\rm int}$. The rotation period of the star with age $t$ is given by
\begin{equation}
    P_{\rm spin}\approx 2  \left({B_{\rm dip}\over 10^{15}}\right)\left({t\over 30\,\hbox{yr}}\right)^{1/2}\hbox{s}.
\end{equation}
The strong $B_{\rm int}$ ensures that the magnetar precesses as a  rigid body \citep{2004ApJ...613.1157L}. 
The period of precession is given by
\begin{eqnarray}
\label{eq:Ppr}
    P_{\rm pr}&\approx&  \frac{P_{\rm spin}}{\epsilon}\\
     &\approx& 20\, k_{0.01}^{-1}\left({B_{\rm int}\over 10^{16}\,\hbox{G}}\right)^{-2}{B_{\rm dip}\over 10^{15}\,\hbox{G}}\left({t\over 30\hbox{yr}}\right)^{1/2}\hbox{d}.\nonumber 
\end{eqnarray}
The magnetic field $B_{\rm dip}\sim 0.1B_{\rm int}\sim 10^{15}$~G is similar to that assumed in the shock maser model of FRBs \citep{2019arXiv190807743B}.\footnote{ Similar fields are  found in the  magnetars observed in our galaxy \citep{2017ARA&A..55..261K}, which are substantially older.
There is also some evidence that the fields in the galactic magnetars are decaying on the timescale comparable to their age (e.g.  \citealp{2019MNRAS.487.1426B}).
}
These values of $B_{\rm int}$ and $B_{\rm dip}$ require $k\sim 0.01$ in order to match $P_{\rm pr}$ with the observed 16-day period. $B_{\rm dip}\sim 0.1 B_{\rm int}$ and $k\sim 10^{-2}$ are both consistent with the magnetic field being tangled inside the magnetar. Alternatively, if the field configuration was simple, then a smaller $B_{\rm int}$ and/or a greater $B_{\rm dip}$ could bring the precession period to agreement with observations. However, these $B_{\rm dip}$ and $B_{\rm int}$  would be in tension with the \cite{2019arXiv190807743B} model of the radio bursts. 

\begin{figure}[t]
\centering
\includegraphics[width=.48\textwidth]{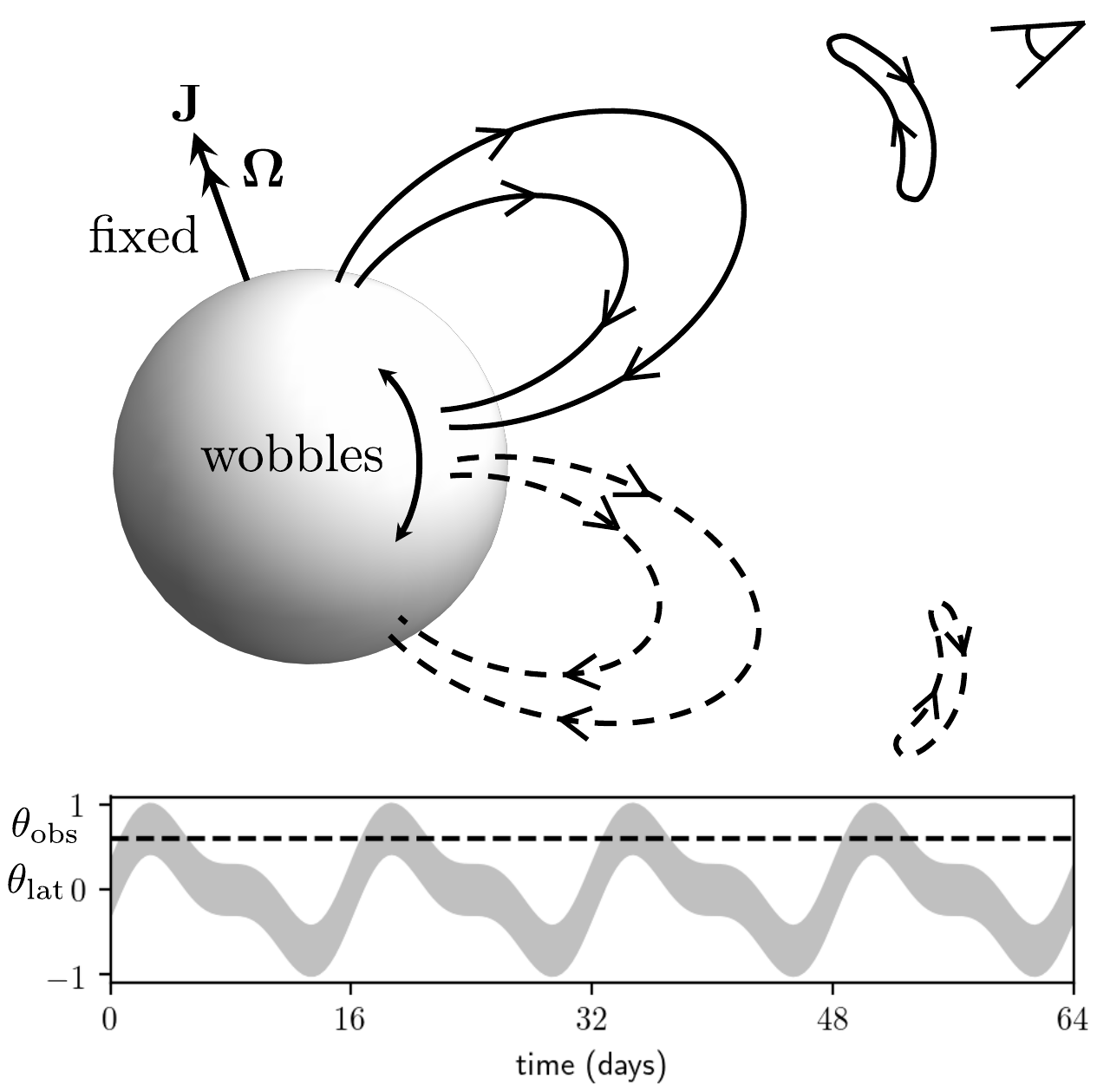}
\caption{Schematic picture of a precessing flaring magnetar, viewed in the fixed lab frame. The angular velocity $\vec{\Omega}$ remains aligned (within a small angle $\sim\epsilon$) with the angular momentum $\vec{J}$, which is conserved and thus unchanging. The flares occur in the active magnetospheric loops attached to the star, which wobbles with a large amplitude and the period $P_\text{pr}$ given in Equation~(\ref{eq:Ppr}). The repeating flares eject relativistic plasmoids, which occupy a limited solid angle and soon take the shape of thin pancakes flying away from the star and driving a blast wave into the magnetar wind (Beloborodov 2019).  The latitudes occupied by the flare ejecta are the directions of the beamed FRB emission from the blast wave. The emission lattitudes are shown in the inserted lower panel by the shaded grey stripe, which periodically intersects the observer line of site (dashed line).
}
\label{figure_1}
\end{figure}
Generally, precession of a triaxial body is not periodic in the laboratory frame. However, the motion of the angular velocity vector in the frame of reference attached to the body is strictly periodic (e.g., \citealp{1969mech.book.....L}).\footnote{This can be quickly seen as follows. In the body frame, the angular velocity $\vec{\Omega}=(\Omega_x, \Omega_y, \Omega_z)$ lies at the intersection of two ellipsoids, which are defined by the rotational energy conservation $\vec{\Omega}\cdot I\vec{\Omega}=2E_{\rm rot}$ and by angular momentum conservation $(I\vec{\Omega})^2=J^2$, where $I$ is the tensor of inertia. 
The intersection of the two ellipsoids is
a closed curve, and $\vec{\Omega}$ moves cyclically along this curve.} The precession of a slightly deformed (nearly spherical) star is special, because its tensor of inertia is very close to  a multiple of the unit matrix. This implies that $\vec{\Omega}$ is nearly aligned with the strictly conserved angular momentum $\vec{J}$, and 
hence the direction of $\vec{\Omega}$ is nearly fixed, within an angle of order $\epsilon$.  Precession is the wobbling of the star relative to the nearly fixed $\vec{\Omega}$ (Figure~1). This wobbling motion can span a large range of angles, which characterizes the amplitude of precession.

The precession-driven wobble was previously invoked in radio pulsars. Most famously, the periodic variations of the pulse profile in PSR B1828--11 were interpreted as a manifestation of free precession \citep{2000Natur.406..484S, 2001ApJ...556..392L, 2001MNRAS.324..811J}. This interpretation was appealing from an observational point of view, but was problematic theoretically if the neutron star core were superfluid. \cite{1977ApJ...214..251S} and \cite{2003PhRvL..91j1101L} showed that even a small amount of pinning  of  superfluid vortices inside the star dramatically affects free precession, either by decreasing its period or by rapidly damping its amplitude.\footnote{This is because the superfluid component acts as a gyroscope. If the superfluid vortices are strongly pinned to the nuclei in the crust or to the magnetic fluxtubes in the superconducting core, the gyroscope is rigidly attached to the star and forced to rotate its axis, tracking the precession motion. Then the back-reaction from the gyroscope increases the precession frequency by several orders of magnitude. If the vortex pinning is not perfect, the vortices are dragged past the pinning sites, causing very high levels of dissipation and thus damping the precession.} {In an attempt to solve this conundrum, some  recent work suggested that precession can occur under special circumstances even if the NS contains pinned superfluid \citep{2019MNRAS.482.3032G}. However, last year’s study by \cite{2019MNRAS.485.3230S} definitively showed that the periodic pulse shape variation in PSR B1828-11 are due to the mode-switching in the pulse shape and not due to precession. As far as we are aware, the current  radio-pulsar data is consistent with Shaham's picture that supefuidity suppresses free precession.}

Hyper-active magnetars are 
likely hot enough to quench neutron superfluidity in the core, as discussed below. Therefore, their precession may be strong.

\subsection{Temperature of the star}

Let $E$ be the magnetic energy of the star and $t$ be the characteristic timescale of ambipolar diffusion, so that the core is heated with rate $L\sim E/t$. A young hyper-active magnetar has $E\sim 10^{49} B_{16}^2$~erg and $t\sim 10^9$~s
\citep{2016ApJ...833..261B} which corresponds to $L\sim 10^{40}E_{49}\,t_9^{-1}$~erg/s. 

The star is cooled by neutrino emission through urca reactions. The direct urca
cooling occurs in neutron stars with masses $M>M_D$. The mass $M_D$ depends on the equation of state of the core matter 
\citep{1998PhRvC..58.1804A,2011PhRvC..84f2802C,2013A&A...560A..48P}
and can significantly exceed the canonical neutron star mass $M=1.4M_\sun$.
A magnetar with mass $M<M_D$ is cooled by modified Urca reactions, which involve a spectator nucleon taking the excess momentum.
Then the cooling rate is given by
\citep{1979ApJ...232..541F}
\beq
\label{eq:Murca}
  \dot{q}_{\nu}^M\sim  7\times 10^{20}
   \,T_9^8 \left(\frac{\rho}{\rho_{\rm nuc}}\right)^{2/3} 
    {\rm ~erg~s}^{-1}{\rm ~cm}^{-3},
\eeq
where $\rho_{\rm nuc}=2.8\times 10^{14}$~g~cm$^{-3}$ is the nuclear saturation density. This expression for the cooling rate is valid if 
the core is made of normal matter, not superfluid or superconducting. Superconductivity is suppressed by the ultrastrong magnetic fields under consideration, $B\sim 10^{16}$~G. 
The onset of neutron superfluidity is theoretically expected at a temperature $T_{\rm crit}\sim 10^8-10^9$~K (see e.g. Figure~5 in 
\citealp{2015SSRv..191..239P}),
which is likely below the core temperature found from the  balance between heating and cooling, 
\beq
   T\approx 9\times 10^8 L_{40}^{1/8}\, {\rm K}.
   \label{temperature}
\eeq

\subsection{Damping of precession by viscosity}

Viscosity of the star tends to damp precession. Free energy available for dissipation in the precessing state is
\beq
  E_{\rm pr}\sim \epsilon {\cal I} \Omega^2.
  \label{Epr}
\eeq
In the minimum energy state the longer axis of the deformed star is perpendicular to $\boldsymbol{\Omega}$. Evolution toward this state occurs because the deformed star periodically (with the precession period) changes its orientation with respect to $\boldsymbol{\Omega}$ by a large angle, and angular momentum conservation implies a periodic perturbation $\delta\Omega\sim \epsilon \Omega$. This causes variation of centrifugal acceleration in the star, inducing variations in deformation and density, 
\beq
   \delta\epsilon\sim \frac{\delta\rho}{\rho} \sim \frac{\Omega\delta\Omega}{G\rho_c}, 
   \label{deltarho}
\eeq 
where $\rho_c\sim 10^{15}$~g/cm$^3$ is the central density\footnote{ \cite{2016MNRAS.458.1660L} computed the damping by considering secondary flows, originally found in a different context by \cite{1972MNRAS.156..419M}. We note that in our situation the Alfv\'en-crossing timescale is smaller than the precession period by a factor of $\sim 10^7$, and the tangled magnetic field is likely anchored in the crust. Therefore, the secondary flows are completely suppressed.}. The density perturbation generates a deviation from the chemical equilibrium, which is damped by the urca reaction of neutrino emission. This process determines the bulk viscosity coefficient $\zeta$, which strongly dominates over the shear viscosity \citep{1989PhRvD..39.3804S}. 

Regardless of $\zeta$, the following general argument 
demonstrates that    bulk 
viscosity is unable to damp precession in FRB 180916.J0158+65.
The energy dissipated during one precession period is given by
\begin{equation}
    \delta E_{\rm diss}=\int_0^{P_{\rm pr}} dt\int \delta P{d\over dt}\left({\delta\rho\over \rho}\right) dV,
    \label{ediss}
\end{equation}
where $\delta P$ is the pressure perturbation\footnote{{This can be seen as follows: the incremental mechanical work done on the stellar material is  $\int P 
(\delta\rho/\rho)dV $ and the corresponding mechanical power is 
$\int P [d(\delta\rho/\rho)/dt] dV$ where $P=P_0+\delta P$ is the pressure and $P_0$ is its unperturbed value. After integrating over a full cycle, the term with $P_0$ drops out and one obtains Eq.~(\ref{ediss}).}}. 
For small damping, $\delta P$ and $d(\delta\rho)/ dt$ are nearly out of phase. 
An upper bound on $\delta E_{\rm diss}$ 
is obtained by assuming that $\delta P$ and $d(\delta\rho)/dt$ are perfectly in phase. Replacing time and volume integrations with multiplications by $P_{\rm pr}$ and $V$, we estimate
\begin{equation}
    \delta E_{\rm max}\sim E_g \left({\delta\rho\over \rho}\right)_{\rm max}^2,
\end{equation}
where $E_g\sim GM^2/R$ is the gravitational energy of the star and we have used $\delta P/P\sim \delta\rho/\rho$. Then from Eqs.~(\ref{deltarho}) and (\ref{Epr}), we obtain
\begin{equation}
    {E_{\rm pr}\over \delta E_{\rm max}}\sim {1\over \epsilon}{GM\over R^3 \Omega^2}\sim 10^{10}\left({10^{-6}\over\epsilon}\right)\left({\hbox{rad}/\hbox{s}\over \Omega}\right)^{2}.
\end{equation}
Since the age of the magnetar is several hundred precession periods, clearly the precession cannot be damped by bulk viscosity.

\subsection{Excitation of precession}

In equilibrium, the angular velocity vector is aligned with the principal axis that has the largest moment of inertia, as this minimizes the rotational energy for a fixed angular momentum. 
It is unclear to us whether the magnetar should be born in the equilibrium state. 
However, we conservatively assume that it does, and explore whether precession can be naturally excited afterwards. A small deflection of the angular velocity vector $\vec{\Omega}$ from this principal axis (hereafter designated as the $z$-axis) results in a small-angle free precession. In what follows we view the dynamics of the precession in the frame of reference attached to the rotating star, with the principal axes of inertia serving as our coordinate axes. In this frame of reference, the precession is seen as the rotation of the angular velocity vector around the $z$-axis with the frequency $\Omega_{\rm pr}=2\pi/P_{\rm pr}$.

{\cite{1970ApJ...160L..11G} examined the evolution of the amplitude of  free precession of a neutron star which is being spun down by an external torque. In Goldreich's computation, the neutron star is assumed to be axially symmetric, and the angular velocity vector $\vec{\Omega}$ is precessing around the symmetry axis $z$ which is also one of the principal axes of inertia. The angle 
$\theta$ between $\vec{\Omega}$ and $z$ is the amplitude of precession. The quantities 
$\theta$ 
and $\Omega$ evolve with time on a similar timescale, 
and their changes are related by
the following equation [cf.~Equations (6) and (7) of Goldreich (1970)]:
\begin{equation}
    {d\log (\sin \theta)\over d\log \Omega}={\cos^2\theta\left(1-{3\over 2}\sin^2\chi\right)\over \sin^2\chi+\sin^2\theta\left(1-{3\over 2}\sin^2\chi\right)}.
    \label{angle}
\end{equation}
Here it is assumed that the star is being spun down by a radiation-reaction torque acting on a rotating magnetic dipole, with $\chi$ being the fixed angle between the dipole axis and $z$. Remarkably, the ellipticity plays no role so long as the precession period is much shorter than the spin-down timescale (which is an approximation used in the derivation of Equation~(\ref{angle})). 
This idealized model may be used as an
order-of-magnitude estimate for 
precession of stars with
more realistic torques and tensors of inertia.

As noted by Goldreich, for $\sin\chi>2/3$ (i.e., for $55^\degree<\chi<125^\degree$), the precession amplitude increases as $\Omega$
decreases. 
Stars with
tangled internal 
fields
with a substantial toroidal component 
have
no a-priori reason for the 
magnetic dipole moment to be aligned with
the $z$-axis.
It is natural to expect that in a significant fraction of cases 
$\chi$ is large and so
the amplitude of precession increases with time. 
For instance, if the magnetar was spun down from an initial period of $\sim 10\,$ms to $P_{\rm spin}\sim 1\,$s, its initially small amplitude of precession $\theta$ could have increased by a factor $\sim 100$. The exact 
increase depends on the configuration
of the magnetosphere.}

{This mechanism of pumping a large precession amplitude requires a seed $\theta\neq 0$. It can be seeded by small kicks of $\theta$ that result from}
sudden changes 
in the direction of the 
angular momentum $\boldsymbol{J}$ or 
changes in the inertia tensor of the dynamic, flaring magnetar
\citep{2000ApJ...543..340T}. Note that a change $\delta\Omega/\Omega\sim 10^{-4}$ was associated with the August 1998 giant flare of the galactic magnetar SGR 1900+14 \citep{1999ApJ...524L..55W}. The timescale of the change was not measured, because the spin period observations had an 80-day gap. 

Let us first consider the kicks in angular momentum. 
Simulations of magnetic flares in axisymmetry suggest a sudden increase in the spin period (Parfrey et al.~2013). 
The direction of $\boldsymbol{\Omega}$ remained unchanged in the axisymmetric simulations, and thus they do not inform us directly about the excitation of precession. However,
real non-axisymmetric flares may well be accompanied by an angular momentum kick that is not aligned with the rotation axis. 
Note that the  duration of the main peak of observed giant flares $\delta t\sim 0.3$~s is shorter than the rotation period. Assuming that the flare ejecta of energy $E_{\rm ej}$ is launched from the twisted magnetosphere not exactly radially but with some impact parameter $b$ comparable to the star's radius, one can estimate the ejected angular momentum as $b E_{\rm ej}/c$. The direction of the lost angular momentum $\delta\boldsymbol{J}$ is determined by the geometry of the flaring magnetosphere and can occur at any angle with respect to $\boldsymbol{J}$.

The presence of $\delta\boldsymbol{J}_\perp$ (perpendicular to $\boldsymbol{J}$) leads to a sudden change in the angle $\theta$ between $\vec{\Omega}$ and the $z$-axis (the principle axis of inertia),
\begin{equation}
    \delta\theta\sim 10^{-7}E_{{\rm ej}, 43}{\cal I}^{-1}_{45}\left({b\over 10\,\hbox{km}}\right)\left({P_{\rm spin}\over 2\,\hbox{s}}\right)\hbox{rad}.
\end{equation}
When viewed in the frame co-rotating with the star and its magnetosphere, the directions of  $\delta\boldsymbol{J}$ are likely correlated over many subsequent flares, as changing the  structure of the magnetosphere with energy $\sim 10^{47}-10^{48}$~erg likely requires many flares. Therefore $\delta\theta$ add coherently for $\sim 1/2$ of the precession period, and their sign changes for the other half. The rate of the flares strongly varies on the precession timescale (see Figure 2 of Amiri et al.~2020), and this justifies treating contributions to $\theta$ from different precessional half-periods as steps in a random walk. 
Therefore, the accumulated impact on $\theta$ from $N_{\rm f}$ flares after time $t$ may be estimated as
\begin{equation}
   \theta_{\rm kicks}\sim \delta\theta N_{\rm f}\sqrt{P_{\rm prec}\over 2t}.
\end{equation}
{For a numerical estimate, let us assume that the magnetar in FRB 180916.J0158+65 has been flaring for $t=10\,$yrs. The number of flares  during this period could be estimated from the fact that $28$ bursts have been observed over $\sim 1\,$yr, with the duty cycle of $\sim 1/4$. During each day the source is visible for only $\sim 1\,$hour. Together, this gives $N_f\sim 3\times 10^4$, and  }
\begin{equation}
    \theta_{\rm kicks}
    \sim 10^{-4} E_{{\rm ej},43}\,{\rm rad}.
    \label{kicks}
\end{equation}
{This rough estimate has a significant uncertainty, because of the large uncertainties in $N_{\rm f}$, $E_{\rm ej}$ and the impact parameter $b$. The latter may be investigated using $3$-dimensional simulations of the magnetospheric flares. Still, even with 
a
$100$-fold amplification 
of $\theta$
due to 
spindown, the angular-momentum kicks do not provide  
a robust
mechanism for 
seeding
large-amplitude precession.}

Generating $\theta$ by rapid movements of the principal axes can be much more efficient.
Such movements happen during the rearrangement of magnetic stresses inside and outside the neutron star.  \cite{1995MNRAS.275..255T} raised the possibility of a large-scale magnetic instability inside a magnetar. A single such event over the lifetime of a magnetar could shift the principal axes of the magnetically deformed star by an angle $\theta\sim 1$. Since the instability occurs on a short (Alfv\'en-crossing) timescale, it instantly excites the large-angle precession.
Alternatively, the principal axes could receive small kicks in many flares of smaller energy. Let $t_I$ be the timescale to accumulate the net shift of axes by $\sim \pi/2$. For a complex evolution of the tangled field,  $t_I<t$ is possible.  A magnetar flaring with a rate $\dot{N}_{\rm f}$ changes the axes in each flare by 
\begin{equation}
    \delta\theta_{\rm mag}\sim 
(\pi/2) (t_I\dot{N}_{\rm flares})^{-1}.
\end{equation}
Just like in the case of angular-momentum kicks, these changes in $\theta$ are correlated over  $\sim 1/2$ of the precessional period, and thus the overall change is given by
\begin{equation}
    \theta_{\rm mag}\sim \frac{\pi}{2}\,\sqrt{P_{\rm prec}\over 2t}{t\over t_I}\sim 0.1\left(\frac{t}{t_I}\right){\rm rad}.
\end{equation}

We conclude that it is possible and perhaps natural for giant flares to stochastically excite free precession with an amplitude of $\simgt 0.1\,$rad, especially if it is aided by the subsequent amplification due to the spin-down. Alternatively, a single large-scale rearrangement of the internal field could excite a large-angle free precession.

\section{Discussion}

We  emphasize that precession as a possible origin of the periodicity in repeating FRB 80916.J0158+65
was first suggested in the discovery paper. The purpose of this Letter is to show that free precession of a magnetar with internal fields $B\sim 10^{16}$~G is indeed capable of economically explaining the FRB observations, with no need of a companion. We find that the expected ellipticity and  spin period of the magnetar give the precession period 
comparable to
the observed 16 day period. Furthermore, flares and internal field rearrangements can excite a significant amplitude of precession, and its damping time is orders of magnitude longer than the age of the magnetar. 

{Several weeks after this paper was submitted, \cite{2020arXiv200303596R} published some evidence for $159$-day periodicity in the first-detected repeating FRB 121102. This period can be easily accommodated within the magnetar precession model, by e.g., assuming that the internal field is $\sim 3$ times smaller than the one in FRB 180916.J0158+65}

Theorists are often blamed for `postdictions’ and it is certainly a fair criticism with regard to this Letter. In fact, historically there is no shortage of theoretical attempts to
predict signatures of precession in magnetars, starting with  \cite{1999ApJ...519L..77M} and \cite{2000ApJ...543..340T}, and yet to date no precession has been observed in galactic magnetars.
This is explained by the presence of a substantial amount of neutron superfluid, 
which
can be inferred  from
the 
observed glitches in  magnetar spin rates
\citep{2014ApJ...784...37D}. The superfluid suppresses free precession via the \cite{1977ApJ...214..251S} mechanism, as a result of strong interaction between the superfluid vortices and the rest of the star.

Superfluidity is less likely in the young hyperactive magnetars proposed as the engines of repeating FRBs, because they are heated with higher rates. Their internal temperatures are capable of reaching $10^9\,$K (Equation~\ref{temperature}), which can be just enough to exceed the critical temperature for superfluidity, $T_{\rm crit}$. Most theoretical estimates give $T_{\rm crit}<10^9K$ (see, e.g., Potekhin et al.~2015). 
It is also consistent with observations of neutron star cooling in the Cassiopea A supernova remnant, which were used to estimate
$5\times 10^8<T_{\rm crit}<9\times 10^8$
\citep{2011MNRAS.412L.108S, 2011PhRvL.106h1101P}. 
 
Precession also requires that the magnetar be not too massive, so that it cools by the modified urca reactions. The much stronger direct urca cooling would be enabled in a massive neutron star, $M > M_D$. It would reduce $T$ below $T_{\rm crit}$ and suppress precession. The condition $M < M_D$ gives a significant constraint on $M$, with the exact upper limit $M_D$ depending on the equation of state of the deep core.

{One  testable prediction of the precession model is that the 
observed period $P_{\rm prec}$
should increase with time 
as the star spins down, according to Eq.~(\ref{eq:Ppr}). As was pointed out to us by Andrei Gruzinov, after time $\Delta t$ the period increase should lead to the fractional phase residual
\begin{equation}
    {\delta t\over P}\sim {\Delta t^2\over 2 t P}\sim 0.4\Delta t_{\rm yr}^2 t^{-1}_{\rm 30yr}.
\end{equation}
While more than half of it can be fitted out by adjusting appropriately the period and the phase of the precession, it is clear that a very constraining measurement is possible within a year. A similar argument was made in \cite{2019arXiv191200526K} as a comment on the first version of this paper. 

Shortly after this paper was submitted, an independent study by \cite{2020arXiv200205752Z} appeared on the arxiv. These authors 
also explore free precession as a mechanism of 16 day periodicity in FRB~180916.J0158+65, and extend their analysis by adopting a specific shape for the angular distribution of the beamed FRB source. This allows them to design a model predicting the distribution of the burst arrival times.
We foresee that this type of modeling may be useful in future for interpreting the timing features of the precession model.  Zanazzi \& Lai  did not address the damping and excitation of the free precession in FRB 180916.J0158+65, which was an important focus of our work.}

We thank Andrei Gruzinov for pointing out to us that the increase in the precession period due to the spindown of
the magnetar is a measurable prediction of our model. We thank Dongzi Li for patiently explaining to us the systematics of the CHIME measurements. We thank Noemie Globus, Elias Most, and Sasha Philippov for useful discussions, and Sarah Levin for help with the prose. AMB is supported by NASA grant
NNX17AK37G, a Simons Investigator Award (grant
$\#$446228), and the Humboldt Foundation.

\bibliography{ms}

\clearpage

\end{document}